\documentclass[12pt,a4paper]{article} 

\usepackage{anysize}
\marginsize{1.5cm}{1.5cm}{1.5cm}{1.5cm}
\usepackage{braket}
\usepackage{mathtools}
\usepackage{babel}
\usepackage{authblk}
\usepackage{amsmath}
\usepackage{amssymb}

\newcommand{\ident}{\mathbb{I}}
\newcommand{\Tr}{\mathop{\mbox{Tr}}\nolimits}

\title{Photon number states via iterated photon addition in a loop}

\author[1]{Barna Mendei}
\author[2]{G\'abor Homa}
\author[2, 3]{P\'eter \'Ad\'am}
\author[2]{M\'aty\'as Koniorczyk}

\affil[1]{Faculty of Natural Sciences, Lor\'and E\"otv\"os University, Budapest, Hungary}
\affil[2]{HUN-REN Wigner Research Centre for Physics, Budapest, Hungary}
\affil[3]{Institute of Physics, University of P\'ecs, P\'ecs, Hungary}

\begin{document}

\maketitle

\begin{abstract}
    We consider the probabilistic generation of time-bin photon number states from a train of single photon pulses. We propose a simple interferometric feedback loop setup having a beam splitter and a possibly non-idel detector. This Hong-Ou-Mandel type scheme implements iterated photon additions. Our detailed study shows that up to 4 photons this simple setup can provide reasonable success probabilities and fidelities.
\end{abstract}


\section{Introduction}

Hong-Ou-Mandel interference~\cite{Hong_1987} is one of the most prominent examples of quantum interference phenomena which is in the heart of many photonic
quantum state engineering and quantum information processing schemata.
When two single photons arrive at each input port of a symmetric beam
splitter at time, they leave the beam splitter as a photon pair,
i.e. a two-photon state in one or the other output ports. The
Hong-Ou-Mandel interference is discussed in full detail in a recent
tutorial by Bra\'nczyk~\cite{1711.00080v1}.

Placing a photodetector to one output port of the beam splitter it is
possible, in principle, to generate a two-photon Fock-state in the
other output. If the detector does not detect any photons, that is, it
realizes a quantum measurement projecting onto the vacuum state, the
other output port is left in the desired two-photon state; a highly
nonclassical and non-Gaussian state. From the quantum
state engineering point of view, this can be understood as adding a
photon to a single-photon state.

Photon addition in general, as a photonic quantum engineering method,
has a broad coverage in the literature, see e.g. the recent review by
Biagi et al~\cite{Biagi_2022}. Realizing photon addition with a beam splitter
has the shortcoming that in the case of non-ideal detectors the
imperfect measurement results in a non-ideal mixed output state. In
heralding schemes, e.g. photon subtraction this a less significant
issue as they rely on the actual detection of a photon. Hence, photon
addition is more often realized using nonlinear optics.  In spite of
that, owing to the significant development of detectors, the
simplicity of using a beam splitter and the fundamental relevance of
the Hong-Ou-Mandel interference, the idea of adding photons to a
few-photon state with beam splitters is not to be ignored.

The generation of photon number states is a topic which has been
discussed broadly in the literature, both theoretically (see
e.g. \cite{PhysRevA.39.2493, PhysRevA.80.013805}) and experimentally,
with various experimental approaches including e.g. cavity
QED~\cite{Varcoe_2000, PhysRevLett.86.3534, PhysRevA.80.013805,
  Sayrin_2011}, micromasers~\cite{Brattke_2003}, interferometric
setups~\cite{Waks_2006}, superconducting quantum
circuits~\cite{Hofheinz_2008}, or quantum
dots~\cite{PhysRevResearch.2.033489}.  Here we consider a less perfect
but rather simple optical interferometric approach.

Recently there have been proposals to generate periodic single-photon
sources~\cite{Adam_2014, PhysRevA.102.013513, MeyerScott2020, arakawa2020progress, adam2023single, li2023quantum, adam2024single}. These would produce a train of photon pulses
with single photon in each of them. In the present work we study a 
loop setup with a single beam splitter,
using a periodic single-photon source as an input, and realizing
iterated conditional photon additions on a Hong-Ou-Mandel
basis. Certainly the schema is theoretically equivalent to the use of
many beam splitters with single-photon inputs. Also, the photon
addition with nonlinear optics could be considered in a similar
setting. However, the use of a single beam splitter and a single
detector is appealing for its simplicity.

We calculate the probability of $n$-photon states in the an ideal
version of the setup. Though this decreases exponentially with $n$,
the probability for a few-photon state is negligible. We also analyze
the impact of having a non-ideal detector to the schema.

\section{Methods}

In the present work we consider time-bin modes of the electromagnetic
field. The pulse shapes are not explicitly taken into account to
maintain the simplicity of the consideration, hence, for each mode we
have a single annihilation operator for each mode so that the
\begin{equation}
    \left[\hat a_i, \hat a_j^\dagger\right]=\delta_{i,j}
\end{equation}
commutation relations hold. (If one adds frequency dependence and
pulse shapes, the interference visibility is described more
efficiently. When the timing of the pulses is appropriate, hence the
visibility is maximal, the same results are obtained as from the
present simiplified analysis.

The considered scheme also contains beam
splitters~\cite{campos1989quantum}. We cosider beam splitters
described by real unitary matrices, whose input and output creation
operators are
\begin{equation}
\label{eq: a_and_c_operators}
  \begin{pmatrix}
    \hat a_1^\dagger \cr
    \hat a_2^\dagger
  \end{pmatrix}
  =
  \begin{pmatrix}
    \sqrt{\tau} & - \sqrt{1-\tau}  \cr
    \sqrt{1-\tau} & \sqrt{\tau}
  \end{pmatrix}
  \begin{pmatrix}
    \hat b_1^\dagger \cr
    \hat b_2^\dagger
  \end{pmatrix},
\end{equation}
where $\tau\in [0,1]$ is the transmittivity of the beam splitter. We
do not consider the phase parameters for the transmitted and reflected
beam as they do not introduce different physics in our considerations.

\begin{figure}
  \centering
  \includegraphics[width=\textwidth]{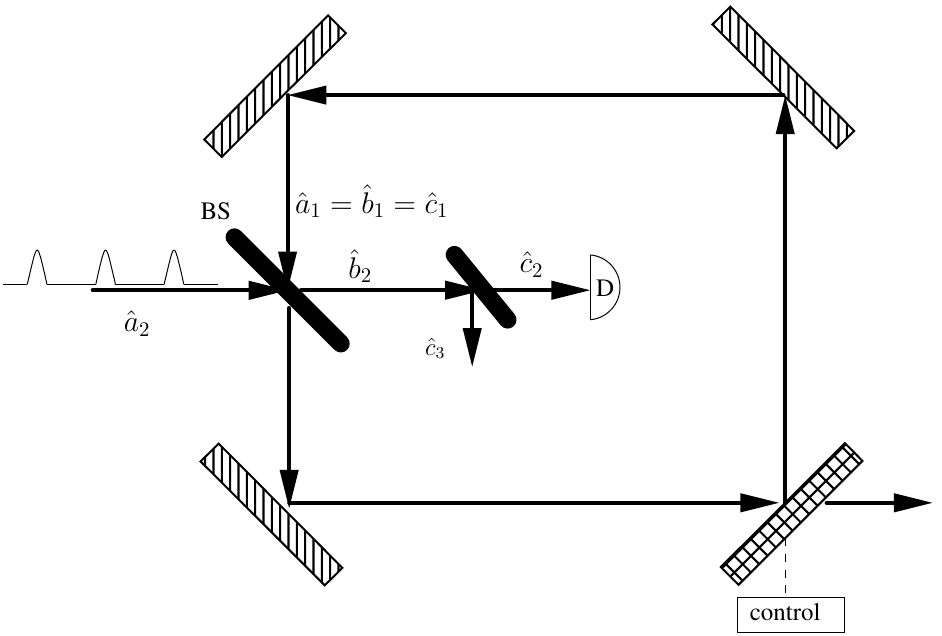}
  \caption{The scheme considered in the present manuscript.}
  \label{fig:schema}
\end{figure}

The particular, the time-multiplexed scheme we consider is depicted in
Fig.~\ref{fig:schema}.
A train of single-photon pulses generated periodically enter the
interferometer in mode $1$. When the first mode arrives, it interferes
with the vacuum in mode $2$. As of the subsequent pulses, the delay loop
formed by the controlled switch completely reflecting the beam and the
two mirrors are set in such a way that each pulse overlaps as exactly
as possible with the previous input pulse. Hence, in each turn, the
state in mode $2$ at the beam splitter input will be equal to the
state of mode $1$ in the previous period.  The other output (mode $2$)
of the beam splitter $BS$ is sent to a threshold detector with
single-photon efficiency. The detector is assumed to be possibly
lossy, with efficiency $\eta$. The detector loss is modeled in the
standard way with an additional beam splitter in front of the detector
with transmittance $\eta$, whose other input mode, mode $3$ is the vacuum and
whose other output mode is ignored (i.e. traced out). Dark counts are
ignored as their rate can be made low in case of threshold detectors.

The way of operating the system is to let a sequence of $n$ pulses
interfere, and couple out the resulting state from the system after
the $n$ pulses. The result is accepted if the detector did not click
during the operation, hence, in each period, vacuum was detected in
mode 2. Under such circumstances, the output state should ideally be
an $n$-photon Fock-state, alebit with a probability decreasing quickly
with $n$. The operation can be interpreted as a photon addition
repeated $n$ times. In what follows we answer the following
questions. What is the exact probability of generating an $n$-photon
Fock state with such a simple scheme? How does this depend on the beam
splitter transmittance $\tau$, and the detector efficiency
$\eta$. What is the actual fidelity of generating an $n$-photon Fock
state, what kind of noise is introduced by the detector loss? Finally,
under what circumstances will the output state still have negative
parts of its Wigner function?

\section{Results}

This section is organized as follows. First we analyze photon addition
with a beam splitter and a detector: the case when a single and an
$n$-photon state interfere on a beam splitter, and one of the modes is
projected to the vacuum state afterwards. This will be a single step
in our iteration.

\subsection{Photon addition with a beam splitter and a detector}

\begin{figure}
    \centering
    \includegraphics[width=0.7\textwidth]{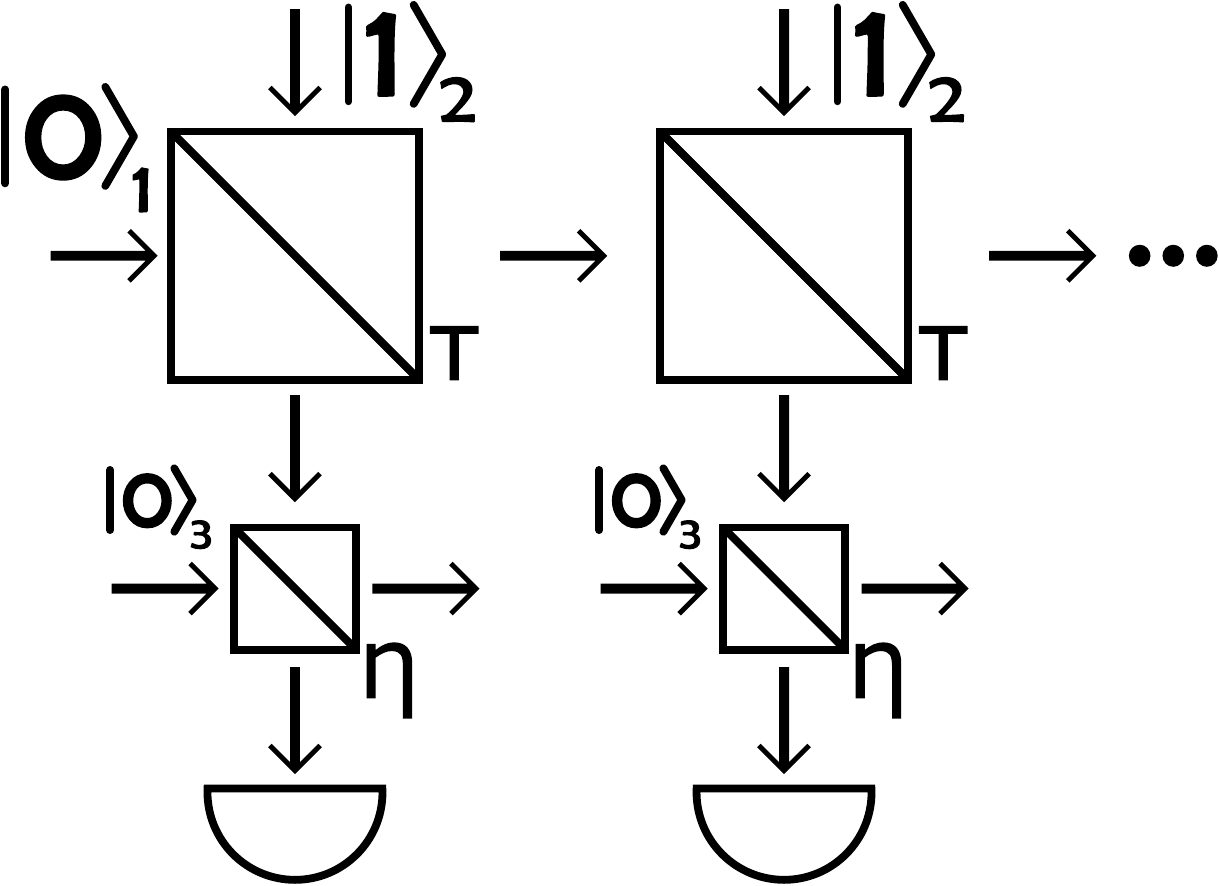}
    \caption{A periodic system of beam splitters with $\tau$ transmittance and real detectors with $\eta$ efficiency. This is equivalent to the loop scheme presented in Figure~\ref{fig:schema}.}
    \label{fig:bs-row}
\end{figure}

Our loop scheme can be replaced by a periodic system consisting of beam splitters and ideal detectors. This can be seen in Figure~\ref{fig:bs-row}. In one period there is the main beam splitter with $\tau$ transmittance and the real detector modelled as described previously (with another beam splitter and an ideal detector). At each step input mode $1$ is the same as output mode except for the first round when it is exactly the vacuum state. Input mode $2$ is where the periodic one-photon states are coming in. Output mode $2$ is lead to the $\eta$ transmittance beam splitter (i.e. the real detector with $\eta$ efficiency), so it is an input mode of it. Its other input mode is mode $3$ where we assume that only vacuum is present, so the ouput mode $2$ is interfering with the vacuum state at the $\eta$ transmittance beam splitter. From it the output mode $2$ is going towards the ideal detector and output mode $3$ is neglected (i.e. traced out).

Let us look at a general period of this system, when $n$ photons are coming onto input mode $1$ $\left(\ket{n}_1\right)$, one photon coming to mode $2$ $\left(\ket{1}_2\right)$, and there is vacuum at mode $3$ $\left(\ket{0}_3\right)$, so the input photon number state is
\begin{equation}
    \ket{\psi_\text{in}}_{123}=\ket{n10}_{123}=\dfrac{1}{\sqrt{n!}}\left(a_1^\dagger\right)^na_2^\dagger\ket{000}_{123},
\end{equation}
where in the last step we expressed it with the bosonic creation operators of each input mode acting on the vacuum state.

First, examine the effect of the first (main) beam splitter with $\tau$ transmittance. It can be described similarly to what we have already presented in (\ref{eq: a_and_c_operators}), but now we have a third input mode, which is not affected by this beam splitter, so its effect on the creation operators is the following
\begin{equation}
\label{eq:bs_btoa}
    \begin{pmatrix}
        \hat a_1^\dagger \cr
        \hat a_2^\dagger \cr
        \hat a_3^\dagger
    \end{pmatrix}=\begin{pmatrix}
        \sqrt{\tau} & -\sqrt{1-\tau} & 0 \cr
        \sqrt{1-\tau} & \sqrt{\tau} & 0 \cr
        0 & 0 & 1
    \end{pmatrix}\begin{pmatrix}
        \hat b_1^\dagger \cr
        \hat b_2^\dagger \cr
        \hat b_3^\dagger
    \end{pmatrix}.
\end{equation}
Using this the output state of the $\tau$ transmittance beam splitter is
\begin{multline}
    \ket{\psi_\text{out}}_{123}=\dfrac{1}{\sqrt{n!}}\left(\sqrt{\tau}\cdot\hat b_1^\dagger-\sqrt{1-\tau}\cdot\hat b_2^\dagger\right)^n\left(\sqrt{1-\tau}\cdot\hat b_1^\dagger+\sqrt{\tau}\cdot\hat b_2^\dagger\right)\ket{000}_{123}=\\
    =(-1)^n\sqrt{\dfrac{\tau(1-\tau)^n}{n!}}\left(\hat b_2^\dagger\right)^{n+1}\ket{000}_{123}+\\
    +\sum\limits_{k=1}^n(-1)^{n-k}\sqrt{\dfrac{\tau^{k-1}(1-\tau)^{n-k}}{n!}}\left[\binom{n}{k}\tau-\binom{n}{k-1}(1-\tau)\right]\left(\hat b_1^\dagger\right)^k\left(\hat b_2^\dagger\right)^{n-k+1}\ket{000}_{123}+\\
    +\sqrt{\dfrac{\tau^n(1-\tau)}{n!}}\left(\hat b_1^\dagger\right)^{n+1}\ket{000}_{123}.
\end{multline}

Now we have to apply the effect of the other beam splitter (standing in front of an ideal detector with $\eta$ transmittance) on this state. It can be expressed similarly:
\begin{equation}
\label{eq:bs_ctob}
    \begin{pmatrix}
        \hat b_1^\dagger \cr
        \hat b_2^\dagger \cr
        \hat b_3^\dagger
    \end{pmatrix}=\begin{pmatrix}
        1 & 0 & 0 \cr
        0 & \sqrt{\eta} & \sqrt{1-\eta} \cr
        0 & -\sqrt{1-\eta} & \sqrt{\eta}
    \end{pmatrix}\begin{pmatrix}
        \hat c_1^\dagger \cr
        \hat c_2^\dagger \cr
        \hat c_3^\dagger
    \end{pmatrix},
\end{equation}
where we introduced $\hat c_i^\dagger$ as the creation operator of the $i$th output mode of this beam splitter. With the proper substitutions we get the following output state:
\begin{multline}
    \ket{\psi_\text{out}'}_{123}=\sum\limits_{k=0}^{n+1}(-1)^n\sqrt{\dfrac{\tau(1-\tau)^n}{n!}}\binom{n+1}{k}\sqrt{\eta^{n-k+1}(1-\eta)^k}\left(\hat b_2^\dagger\right)^{n-k+1}\left(\hat b_3^\dagger\right)^k\ket{000}_{123}+\\
    +\sum\limits_{k=1}^n\sum\limits_{l=0}^{n-k+1}(-1)^{n-k}\sqrt{\dfrac{\tau^{k-1}(1-\tau)^{n-k}}{n!}}\sqrt{\eta^{n-k-l+1}(1-\eta)^l}\times\\
    \times\left[\binom{n}{k}\tau-\binom{n}{k-1}(1-\tau)\right]\binom{n-k+1}{l}\left(\hat b_1^\dagger\right)^k\left(\hat b_2^\dagger\right)^{n-k-l+1}\left(\hat b_3^\dagger\right)^l\ket{000}_{123}+\\
    +\sqrt{\dfrac{\tau^n(1-\tau)}{n!}}\left(\hat b_1^\dagger\right)^{n+1}\ket{000}_{123}=\\
    =\sum\limits_{k=0}^{n+1}(-1)^n\sqrt{\tau(1-\tau)^n\eta^{n-k+1}(1-\eta)^k}\sqrt{\binom{n+1}{k}(n+1)}\ket{0; n-k+1; k}_{123}+\\
    +\sum\limits_{k=1}^n\sum\limits_{l=0}^{n-k+1}(-1)^{n-k}\sqrt{\tau^{k-1}(1-\tau)^{n-k}\eta^{n-k-l+1}(1-\eta)^l}\sqrt{\binom{n-k+1}{l}\dfrac{(n-k+1)!k!}{n!}}\times\\
    \times\left[\binom{n}{k}\tau-\binom{n}{k-1}(1-\tau)\right]\ket{k; n-k-l+1; l}_{123}+\\
    +\sqrt{\tau^n(1-\tau)}\sqrt{n+1}\ket{n+1; 0; 0}_{123}
\end{multline}

We are interested in the cases where the detector does not detect any photons. This state can be obtained by a projection. The operator we can use for this is
\begin{equation}
    \hat \ident^{(1)}\otimes\hat P_0^{(2)}\otimes\hat \ident^{(3)}=\sum\limits_{i=0}^\infty\ket{i}_1\bra{i}_1\otimes\ket{0}_2\bra{0}_2\otimes\sum\limits_{j=0}^\infty\ket{j}_3\bra{j}_3,
\end{equation}
where $\ident^{(i)}$ is the identity of the $i$th mode and $P_k^{(i)}=\ket{k}_i\bra{k}_i$ is the projector of the $i$th mode projecting to the $k$-photon state. Applying this operator on $\ket{\psi_\text{out}'}_{123}$ we get
\begin{multline}
    \hat \ident^{(1)}\otimes\hat P_0^{(2)}\otimes\hat \ident^{(3)}\ket{\psi_\text{out}'}_{123}=\\
    =(-1)^n\sqrt{\tau(1-\tau)^n(1-\eta)^{n+1}}\sqrt{n+1}\ket{0;0; n+1}_{123}+\\
    +\sum\limits_{k=1}^n(-1)^{n-k}\sqrt{\tau^{k-1}(1-\tau)^{n-k}(1-\eta)^{n-k+1}}\sqrt{\dfrac{(n-k+1)!k!}{n!}}\times\\
    \times\left[\binom{n}{k}\tau-\binom{n}{k-1}(1-\tau)\right]\ket{k; 0; n-k+1}_{123}+\sqrt{\tau^n(1-\tau)}\sqrt{n+1}\ket{n+1; 0; 0}_{123}.
\end{multline}
Its absolute value squared is the probability of not detecting any photon with the detector. It is
\begin{equation}
    p=\left(\eta^2\tau\left(1-\tau\right)\left(n+1\right)+1-\eta\right)\left(\eta\tau+1-\eta\right)^{n-1}.
\end{equation}

Consider the basic Hong--Ou--Mandel setup with $\tau=0.5$ as an example and an ideal detector $\left(\eta=1\right)$. In this case $p$ decreases exponentially\footnote{We note, that usually, apart from some special cases, $p$ decreases exponentially.}:
\begin{equation}
    p(n, \tau=0.5, \eta=1)=\left(n+1\right)2^{-n-1}.
\end{equation}

Back to the general case, after the projection the state we got is not normalised, that is why, we have to multiply it with $1/\sqrt{p}$ to get a properly normalised photon state:
\begin{multline}
    \ket{\psi_\text{out, norm}}_{13}=\dfrac{1}{\sqrt{p}}\left\{(-1)^n\sqrt{\tau(1-\tau)^n(1-\eta)^{n+1}}\sqrt{n+1}\ket{0; n+1}_{13}\right.+\\
    +\sum\limits_{k=1}^n(-1)^{n-k}\sqrt{\tau^{k-1}(1-\tau)^{n-k}(1-\eta)^{n-k+1}}\sqrt{\dfrac{(n-k+1)!k!}{n!}}\times\\
    \left.\times\left[\binom{n}{k}\tau-\binom{n}{k-1}(1-\tau)\right]\ket{k; n-k+1}_{13}+\sqrt{\tau^n(1-\tau)}\sqrt{n+1}\ket{n+1; 0}_{13}\right\}.
\end{multline}
Here we no longer denote the photon number in mode $2$, because it is $0$ in every component. Now with the two remaining modes $1$ and $3$ we can obtain the density matrix as follows:
\begin{equation}
    \hat \rho=\ket{\psi_\text{out, norm}}_{13}\bra{\psi_\text{out, norm}}_{13}.
\end{equation}
As we described earlier output mode 3 is ignored, so we have to trace out in it:
\begin{multline}
    \hat \rho_1=\Tr_3\hat{\rho}=\dfrac{1}{p}\left\{\tau(1-\tau)^n(1-\eta)^{n+1}(n+1)\ket{0}_1\bra{0}_1+\right.\\
    +\sum\limits_{k=1}^n\tau^{k-1}(1-\tau)^{n-k}(1-\eta)^{n-k+1}\dfrac{(n-k+1)!k!}{n!}\left[\binom{n}{k}\tau-\binom{n}{k-1}(1-\tau)\right]^2\ket{k}_1\bra{k}_1+\\
    \left.+\tau^n(1-\tau)(n+1)\ket{n+1}_1\bra{n+1}_1\right\}.
\end{multline}
From that fidelity can be obtained as the coefficient of the $\ket{n+1}_1\bra{n+1}_1$ term
\begin{equation}
    F(n, \tau, \eta)=\dfrac{\tau^n(1-\tau)(n+1)}{p(n, \tau, \eta)}.
\end{equation}

\subsection{Iterated photon addition}

Now let us consider the case when a train of $n$ single-photon pulses
arrive at our arrangement, having beam splitter of transmittance
$\tau$ and a detector of efficiency $\eta$. After the $n$ pulses the
result is coupled out.

As before let us denote the absorption operators of the modes after an
iteration, i.e. at the stage before the next input pulse arrives and
when the detection takes place, after the beam splitter modeling
detector loss, by $\hat c_1$, $\hat c_2$, and $\hat c_3$,
respectively. Before the beam splitter modeling loss we have the operators $\hat b$ given by Eq.~\eqref{eq:bs_ctob},
which are expressed with the absorption operators $\hat a$ at the arrival of the
input pulse of this turn according to Eq.~\eqref{eq:bs_btoa}.
As pointed out before, on the condition that the detector never fires,
at each iteration the state in mode 1 emerging from the previous
iteration, $\rho^{(1,n)}$ which is now the input state in mode 2 of
the next iteration, $n+1$ is an incoherent mixture of photon number
states up to $n$ photons. As pointed out before, $\rho^{(1,n)}$ is
diagonal in the photon number basis.

Thus the output at this iteration can be calculated by taking all
possible $k$-photon inputs from $1$ to $n$ at mode 2 of the beam splitter, calculating
the output states before the detection as
\begin{equation}
  \label{eq:psin3out}
  \ket{\Psi_{3|k}}= \frac{1}{\sqrt{k!}} \hat a_1^\dag \left(\hat a_2^\dag\right)^k \ket{000},
\end{equation}
and mixing them with the respective probablilities, resulting in the
density matrix of the three-mode field before the (ideal) detection:
\begin{equation}
  \varrho^{(n+1, \text{preproj})} = \sum_{k=0}^{n} \rho^{(1,n)}_{k,k} \ket{\Psi_{3|k}}\bra{\Psi_{3|k}}.
\end{equation}
The ideal detection is described then with a projection of this
three-mode density operator to the vacuum state in mode 2, leading to
the unnormalized density operator
\begin{equation}
  \varrho^{(n+1, \text{afterproj})} = \ket{0}_2\bra{0}_2 \varrho^{(n+1, \text{preproj})} \ket{0}_2\bra{0}_2.
\end{equation}
From this the probability of the vacuum detection, i.e. that of
obtaining an approximate n+1 photon state on condition that the
previous iteration succeeded is
\begin{equation}
  \label{eq:pn}
  p_{n+1|n}= \Tr \varrho^{n+1, \text{afterproj}}, 
\end{equation}
whereas the state of mode 1 after this iteration, which will be the
input to the next iteration, or the output if the desired number of
iterations took place, will read
\begin{equation}
  \label{eq:rhonplus1}
\rho^{(1,n+1)}=\frac{1}{p_{n+1|n}} \Tr_{2,3} \varrho^{n+1, \text{afterproj}}.
\end{equation}
As pointed out before, this will remain diagonal.
The net success probability will be the product of success probabilities at each turn,
\begin{equation}
  \label{eq:pnnet}
  p_{n+1}=\prod_{k=0}^{n+1}p_{n+1|n}.
\end{equation}

\begin{figure}
\centering
\includegraphics{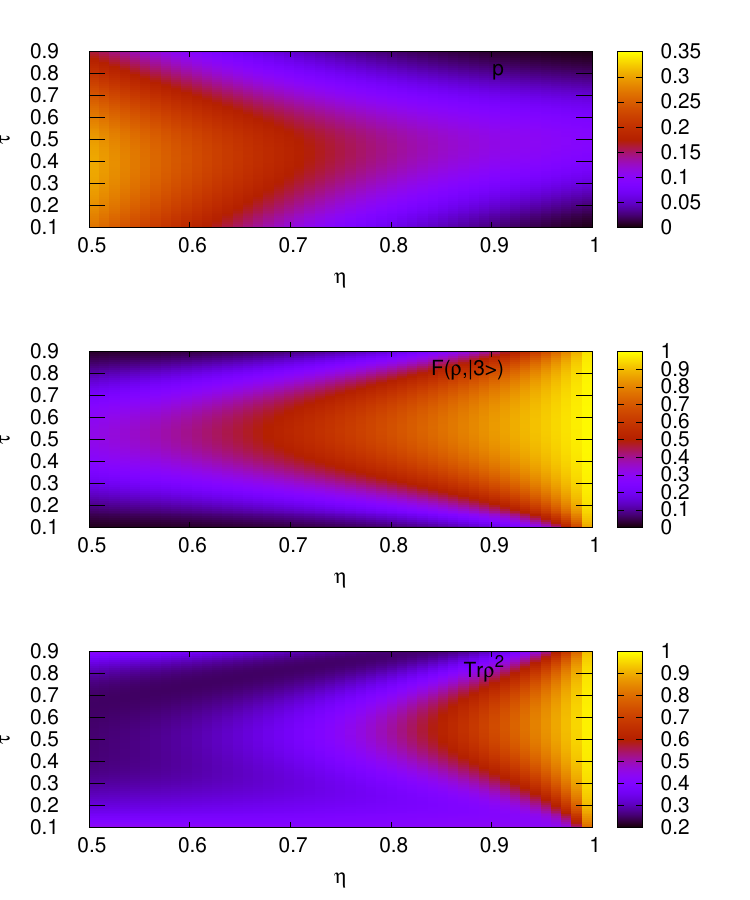}
\caption{Success probability (top), output state fidelity (middle), and output state purity (bottom) as a function of detector efficiency $\eta$ and beam splitter transmittance $\tau$ for generating a state approximating $\ket{3}$ with iterated photon addition.}
\label{fig:3photon}
\end{figure}

\begin{figure}
\centering
\includegraphics{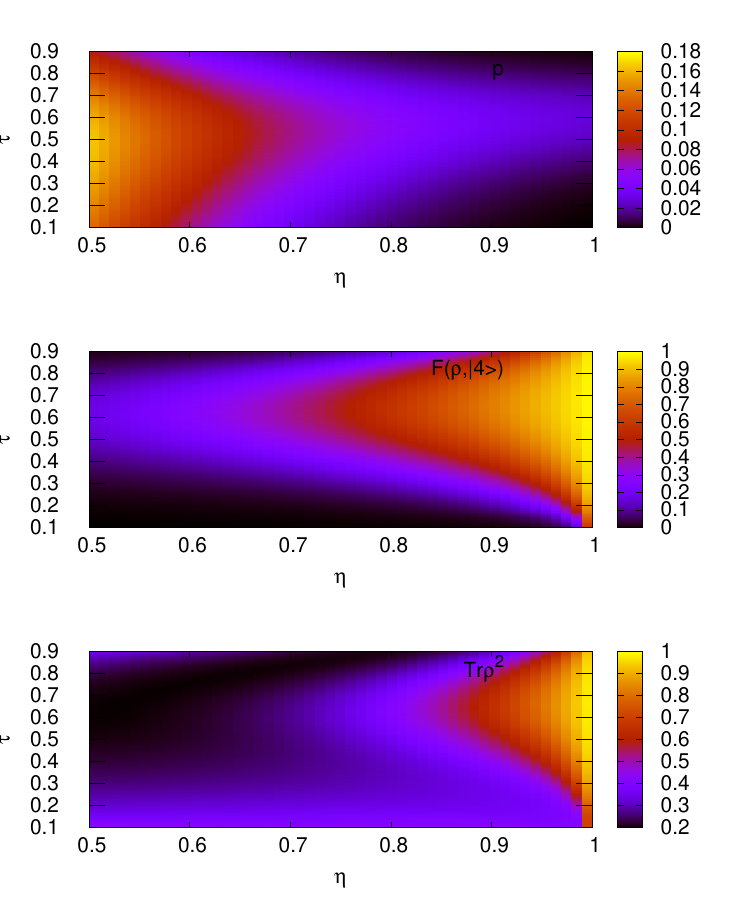}
\caption{Success probability (top), output state fidelity (middle), and output state purity (bottom) as a function of detector efficiency $\eta$ and beam splitter transmittance $\tau$ for generating a state approximating $\ket{4}$ with iterated photon addition.}
\label{fig:4photon}
\end{figure}

These iterations can be evaluated numerically. We have considered the cases of generating 3 or 4 photons in this setup. In Figs.~\ref{fig:3photon} and~\ref{fig:4photon} we have plotted the net success probability, the fidelity of the output state to the desired number state, and the output purity $\Tr \varrho^2$, for the 3 and 4 photon cases, respectively, as a function of detector efficiency $\eta$ and beam splitter transmittance $\tau$.
The figures support the following conclusions. Lower detector efficiencies yield to bigger success probabilities; the detector loss results in a more frequent detection of vacuum. Meanwhile the fidelity and the purity do not decrease too dramatically with the detector efficiency, e.g. in the 3-photon case a detector with an efficiency of $80\%$ will give a fidelity of $0.67$ with a symmetric beam splitter, with a success probability of $0.14$. This is to be compared with the case of the ideal detector where the fidelity is $1$ but the success probability is $0.09$.

\begin{figure}
\centering
\includegraphics[width=0.5\textwidth]{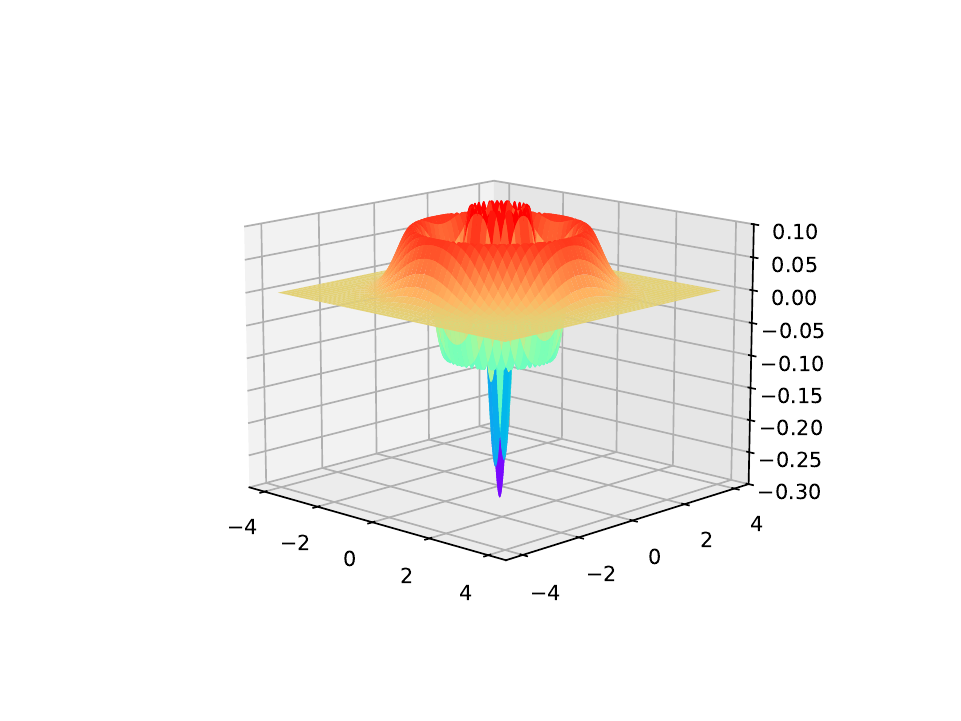}~\includegraphics[width=0.5\textwidth]{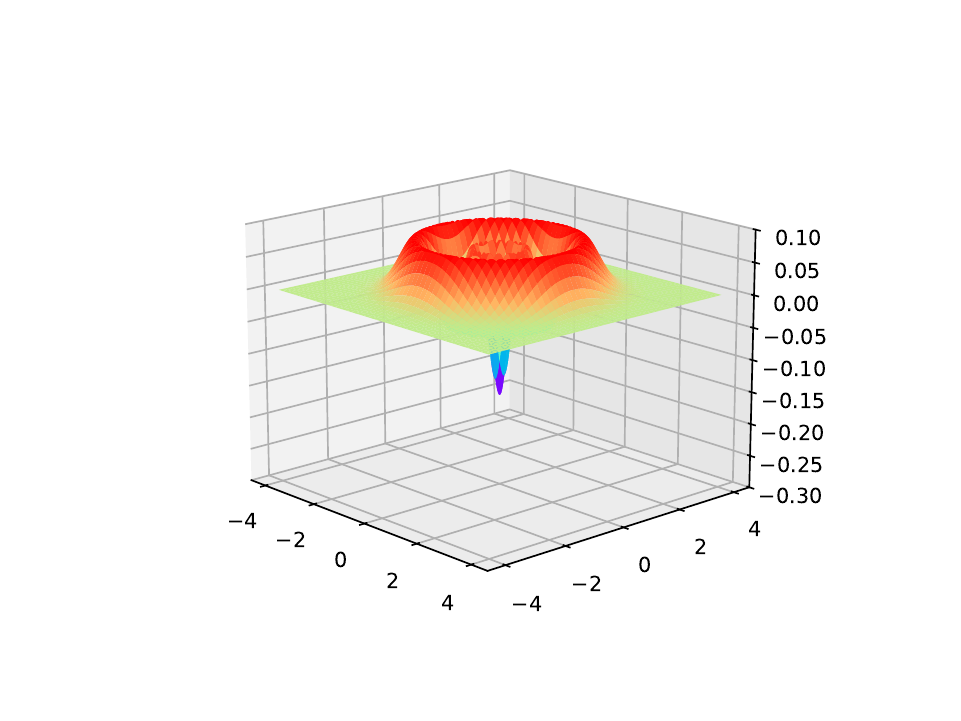}
\caption{Wigner function of the 3-photon Fock state (left) and the one genearated with iterated photon addition with a symmetric beam splitter and a detector with $80\%$ efficiency.}
\label{fig:wigner3}
\end{figure}

While the lower fidelity can be a significant issue in certain applications, we remark that such a state is still highly nonclassical: its Wigner function is very similar in shape to that of the $\ket{3}$ state, with a relevant negative part, as illustrated in Fig.~\ref{fig:wigner3}

Another phenomenon to observe in the figures is that for a given detector efficiency, non-symmetric beam splitters may give a better performance. This is even more marked in the case of the 4-photon state, albeit the results also confirm that the success probability becomes really very low as expected. 

\section{Discussion and conclusions}

We have considered an iterated photon addition scheme using a single
detector and a beam splitter, which probabilistically converts a train
of single-photon pulses into a higher photon number state. Using
nonideal detectors the scheme outputs a mixture of photon states which
can be still close to the desired $n$-photon state, and exhibits
nonclassical properties such as negativity of the Wigner
function. Although the success probability of the scheme decreases,
exponentially with $n$, 3 or 4-photon states can be generated still
with reasonable probability. This probability increases when using an
nonideal detector, however, the fidelity of the generated state to the
desired Fock-state decreases. We have analyzed in detail this
interplay between the success probability and the fidelity. We have
found that in certain cases a better fidelity can be achieved by using
a beam splitter with optimally chosen transmittance.  The presented
scheme thus offers a relatively simple way of generating highly
nonclassical states of time-bin modes. A logical continuaton of the
present research could be the replacement of the beam splitter with a
nonlinear element, and a detailed comparsion with a classical
description of the scheme.

Viewed from a broader context, time multiplexed photon interference
schemes are developing significantly. For instance, in addition to the
time-bin resolved (local) approach studied here has been extended to
time-bucket (global) coincidence, showing nonclassical effects effect
beyond Hong-Ou-Mandel type
interference~\cite{PhysRevLett.125.213604}. In this experimental
context the present schema could easily realized, provided that
periodic single-photon source is available, which latter is also
important in photonic quantum information applications. Let us also
remark that certain optical quantum processors, such as Borealis also
employ delay loops~\cite{Madsen_2022}, hence, the present results can be instructive also
for the detailed understanding of these.

\subsection*{Funding}

This research was supported by the National Research, Development, and Innovation Office of Hungary under the Quantum Information National Laboratory of Hungary (Grant No. 2022-2.1.1-NL-2022-00004), the "Frontline" Research Excellence Program, (Grant. No. KKP 133827), and Project no.~TKP2021-NVA-04. B.M. and M.K. received support from the \'UNKP-23-2 New National Excellence Program of the Ministry for Culture and Innovation from the source of the National Research, Development and Innovation Fund.

\subsection*{Acknowledgments}

The authors thank Zolt\'an Zimbor\'as for his questions that have intitated and motivated the present research, and Aur\'el G\'abris for useful discussions.


\end{document}